\documentclass[a4paper,10pt]{llncs}
\usepackage[utf8]{inputenc}
\usepackage{makeidx}
\usepackage{amsmath}
\usepackage{bbold}
\usepackage{caption}

\newcommand{\ket}[1]{| #1 \rangle}
\newcommand{\bra}[1]{\langle #1 |}
\newcommand{\ketbra}[2]{| #1 \rangle \langle #2 |}
\newcommand{\braket}[2]{\bra{#2}{#1}\rangle}

\newcommand{\1}{\mathbb{1}}

\newcommand{\Complex}{\mathbb{C}}
\usepackage{enumerate}
\usepackage[hidelinks]{hyperref} 
\begin{document}
\title{Quantum image classification using principal component analysis}
\titlerunning{Quantum image classification using PCA}
\toctitle{}

\author{Mateusz Ostaszewski\inst{1,2} \and Przemysław Sadowski\inst{1} \and Piotr Gawron\inst{1}}
\authorrunning{M. Ostaszwski, P. Sadowski, P. Gawron}
\tocauthor{}
\institute{Institute of Theoretical and Applied Informatics, Polish Academy of
Sciences, Ba\l{}tycka 5, 44-100 Gliwice, Poland  \and 
Institute of Mathematics, Silesian University of Technology, 
Kaszubska 23, Gliwice 44-100, Poland
}

\maketitle
\email{mostaszewski@iitis.pl}
\begin{abstract}
We present a novel quantum algorithm for classification of images. The algorithm
is constructed using principal component analysis and von Neuman quantum
measurements. In order to apply the algorithm we present a new quantum
representation of grayscale images.
\end{abstract}
\section{Introduction}
At the end of the last century a new paradigm of computation was proposed
\textit{i.e.} quantum computation. Although it is not yet obvious whether useful
quantum computers can be constructed, the field of quantum algorithms
development in recent years progresses very rapidly \cite{bacon2010recent},
\cite{ambainis2014recent}. For example many new algorithms for quantum machine
learning and quantum image processing were recently created
\cite{lloyd2013quantum}, \cite{schuld2014introduction}.

In this work we introduce an algorithm for image classification of grayscale
images based on classical principal component analysis (PCA) and quantum
measurement. The general idea behind the algorithm is following. Given a set of
training images, using PCA we train a classifier
to detect images similar to those in the training set. Effectively we divide the
image signal space into two orthogonal subspaces. The first one --- spanned by
the leading principal components --- catches the most of the variability of the
signal in the training set, the second one consists mostly of noise.

After the classifier is constructed the leading principal components are used to
create a projector onto a subspace of quantum states. The image which
is being classified is also encoded on a quantum state, and then
measured using the projector defined above. 
%
%

The paper is organised as follows in Section~2 we recall basic notions of
quantum computation, in Section~3 we shortly discuss state of the art in quantum
image processing, in Section~4 we introduce image classification algorithm. And
finally in Section 5 we draw conclusions.
\section{Essentials of quantum computation}
Lets consider the most basic model of a quantum system -- a qubit --
elementary quantum system with two basic physical states. In order to provide 
mathematical description of a state of a qubit we choose an orthonormal 
basis in the 
corresponding Hilbert Space. In 
this case we consider two dimensional Hilbert Space. Our basis will consist two 
vectors that in the braket notation take the form
\begin{equation}
\ket{0}=\left[\begin{array}{c}
1\\
\vdots\\
0
\end{array}\right],\ldots,\  \ket{k}=\left[\begin{array}{c}
0\\
\vdots\\
1
\end{array}\right].
\end{equation}
The $\ket{x}$ vector is called `ket' and its Hermitian conjugation 
$(\ket{x})^\dagger=\bra{x}$ is called `bra'.
We can represent any valid state of a qubit $\ket{\psi}$ as normalized linear 
combination of the 
basis 
vectors:
\begin{equation}
\ket{\psi}=\alpha_1\ket{0}+\cdots +\alpha_k\ket{k},
\end{equation}
where $\alpha_1,\ldots ,\alpha_k\in\mathbb{C}$ and $\sum\limits_{i=1}^{k}|\alpha_i|^2=1$.

The operation which allows us to join $k$ independent qunit systems is the 
tensor 
product. Lets take $k$ qunit
states 
\begin{equation}
\ket{\psi}=\left[\begin{array}{c}
\psi_1\\
\vdots\\
\psi_k
\end{array}\right]=\psi_1\ket{0}+\cdots +\psi_k\ket{k},\ \ket{\phi}=\left[\begin{array}{c}
\phi_1\\
\vdots\\
\phi_k
\end{array}\right]=\phi_1\ket{0}+\cdots +\phi_k\ket{1}.
\end{equation}
We can write their joint state in $\Complex^k\otimes\Complex^k$ as

\begin{equation}
\begin{split}
\ket{\psi}&\otimes\ket{\phi}=
\left[\begin{array}{c}
\psi_1\phi_1\\
\vdots\\
\psi_1\phi_k\\
\psi_2\phi_1\\
\vdots\\
\psi_k\phi_{k-1}\\
\psi_k\phi_k
\end{array}\right].
\end{split}
\end{equation}

The other way of joining quantum systems into a bigger one is use of the direct 
sum. The joint state of two states $\ket{\psi}, 
\ket{\phi}\in\mathbb{C}^k$ 
is
\begin{equation}
\begin{split}
\ket{\psi}\oplus\ket{\phi} =
\left[\begin{array}{c}
\psi_1\\
\vdots\\
\psi_k\\
\phi_1\\
\vdots\\
\phi_k
\end{array}\right].
\end{split}
\end{equation}

We can also consider more general quantum systems called qu$d$its.
Let $\ket{\psi}$ (ket) be a normed column vector from Hilbert space
$\mathbb{C}^n$ with orthonormal basis $\{\ket{i}\}_{i=1}^n$. Dual vector to 
ket
is $\bra{\psi}$ (bra). In such case the state of the system is represented as 
$\ket{\psi}=\sum_i \psi_i 
\ket{i}$.

We denote inner
product of $\ket{\psi}$ and $\ket{\phi}$
by
$\braket{\psi}{\phi}=\sum\limits_{i=1}\limits^{n}\phi_i^*\psi_i.$ 
 It has three properties:
\begin{enumerate}
\item $\braket{\psi}{\psi}\geq 0$ where equality holds iff $\ket{\psi}=0$,
\item $\braket{\psi}{\phi}=\braket{\phi}{\psi}^*$,
\item 
$\bra{\psi}(a_1\ket{\phi_1}+a_2\ket{\phi_2})=a_1\braket{\phi_1}{\psi}+a_2\braket{\phi_2}{\psi}$.
\end{enumerate} 
Furthermore
$\ketbra{\psi}{\phi}=\sum\limits_{i=1}\limits^n\sum\limits_{j=1}\limits^n\psi_i\phi_j^*\ketbra{i}{j}$
 will be their outer product.



One of the most important concepts in quantum information is the measurement. 
The mathematical model of a measurement is as follows. At first we define a
set of outcomes $\Gamma$.  Then we assign corresponding measurement operators 
$\{P_\gamma\}_{\gamma\in\Gamma}$. We request that the measurement operators 
satisfy the condition $P_\gamma^2=P_\gamma$ and $\sum_{\gamma} P_\gamma=\1$.

The probability that we obtain outcome $\gamma$ when measuring a state 
$\ket{\phi}$ is  equal to

\begin{equation}
P_\Gamma(\gamma, \ket{\phi})=\bra{\phi}P_\gamma\ket{\phi}.\label{probab}
\end{equation}
 If we
instantly measure the system for the second time, the outcome willstill be 
equal to
$\gamma$ with certainty because after the first measurement the state of the 
system changes into a state
$$
\frac{P_\gamma\ket{\phi}}{\bra{\phi} P_\gamma\ket{\phi}^{1/2}}.
$$

\section{Quantum image processing --- state of the art}
There are various ways in which classical data can be encoded on quantum states.
The specific encoding depends on the type of the data and quantum algorithms
that one wishes to execute.
\subsection{Quantum representations of digital images}
 Below we recall various representations of quantum
images proposed in recent years.

In the Qubit Lattice representation of grayscale images proposed in~\cite{venegas2003storing}
the intensity of pixel at position $y,x$ is encoded on qubit $\ket{q}_{y,x}$.

The Real Ket representation introduced in~\cite{latorre2005image} stores $2^n\times 2^n$
grayscale images in unnormalised quantum states of the form
$$
\ket{\Psi} = \sum_{i_1,\ldots, i_n=1,\ldots, 4} c_{i_1,\ldots, i_n}\ket{i_1,\ldots, i_n}, 
$$
where $c_{i_1,\ldots, i_n}\in \mathbb{R}$ and subsequent qu$4$its serve as the
position of pixel encoded in a quad-tree.

Flexible representation of quantum images (FRQI) captures information about 
pixel colours and their corresponding position. It is inspired by the pixel
representation for images in the classical computers. The information is gather 
into a quantum state defined as follows
\begin{equation}
| I(\theta)\rangle =
\frac{1}{2^n} \sum_{i=0}^{2^{2n} - 1}
( \cos \theta_i \, |0\rangle + \sin \theta_i \, |1\rangle ) \otimes |i\rangle,
\end{equation}
where $\theta_i \in \left[0 , \frac{\pi}{2} \right]$ and constitutes the vector
encoding colours, $|0 \rangle , |1\rangle$ is a fixed basis of a two dimensional
complex Hilbert space, and $| i \rangle$ is a basis of $2^{2 n}$ dimensional
space responsible for encoding position in the image. The colour is encoded in a
$2D$ vector by $\cos \theta_i \, |0\rangle + \sin \theta_i \, |1\rangle $ which
is connected by a Kronecker product with a vector  $| i \rangle$ responsible for
a position in the image.

A novel enhanced quantum representation (NEQR) of digital images proposed in
\cite{zhang2013neqr} encodes a grayscale $2^n \times 2^n$ image in a quantum
state of the form
$$
\ket{I} = \frac{1}{2^n} \sum\limits_{y=0}^{2^n-1} \sum\limits_{x=0}^{2^n-1} 
\bigotimes_{i=0}^{q-1} \ket{C^{i}_{yx}}\otimes\ket{YX},
$$
where $C^{i}_{yx}$ is a discrete value of image intensity, quantised with $q$
levels of quantisations of pixel at position $(y,x)$.

\subsection{Quantum image processing algorithms}
The sub-field of quantum computation that deals with algorithms development for
quantum image processing is developing very rapidly. At least a hundred papers
discussing this subject were published in recent fifteen years.

It should be noted that, some of classical image transformations already have
their quantum analogues. For example we can list here quantum Fourier
transform~\cite{nielsen2010quantum}, quantum discrete cosine
transform~\cite{klappenecker2001discrete,tseng2005quantum}, and quantum wavelet
transform~\cite{fijany1998quantum}.

There exists several clever techniques to process images encoded in quantum
states for example in \cite{curtis_towards_2004} authors propose a way to
perform template matching algorithm using quantum Fourier transform and
amplitude amplification. In paper~\cite{le2011strategies} the authors extended
the use of quantum circuit models for quantum image representation and
processing. They developed three strategies to extend the number of geometric
transformations~\cite{le2010fast} on quantum images using the FRQI
representation of quantum images. In \cite{yuan2013quantum} authors propose
quantum algorithms for edge detection and image filtering based on projective
measurement. In \cite{yuan_sqr_2014} the authors propose a model for storing and
operating on infra-red images.

Complex quantum image processing requires a number of basic algorithmic
primitives. In~\cite{wang2014image} authors develop quantum image translation,
which maps the position of each picture element into a new position.
In~\cite{zhou2014multidimensional} an algorithm for comparing colour quantum
images based on FRQI model is described.

\section{Algorithm}
The aim of our algorithm is classification of quantum images. 
The input of the algorithm is a quantum representation of an image which we want to test. Algorithm requires a set of principal components. The output is ``$\mathrm{yes}$'' or ``$\mathrm{no}$'' and answers the question whether the image exhibits features represented by principal components.
\subsection{Principal Component Analysis}
In order to create our quantum classifier we use Principal Component Analysis (PCA). This technique has been
successfully applied in the domain of signal processing to various datasets. In celebrated classical paper \cite{turk1991face} it was applied to classification of human faces.

Suppose we have matrix of data $A\in M_{m,n}$ with rank $k\leq m$. The matrix is
composed of vertically stacked horizontal sample vectors. We assume that our
samples are normalised \textit{i.e.} have $l_2$ norm equal to one.

Then by SVD we have
$
A=U\Sigma V^T.
$
where $U\in M_m$ and $V\in M_n$ are orthogonal matrices. 
The matrix $\Sigma = \mathrm{diag}\{\sigma_1,\ldots ,\sigma_q\}$ is such that
$
\sigma_{1}\geq\sigma_{2}\geq \ldots \geq\sigma_{k}>\sigma_{k+1}
= \ldots = \sigma_{q} = 0,
$
with $q=\min(m,n)$.

The numbers $\sigma_{i}$ are called \emph{singular values},
\textit{i.e.} non-negative square roots of the eigenvalues of $AA^T$. The columns of $U$
are eigenvectors of $AA^T$ and the columns of $V$ are eigenvectors of $A^T A$.
The $i$-th column vector of the matrix $U_{:,i}$ is called the $i$-th principal
component of the data.

\subsection{Quantum image representation}

Suppose we have a features vector of $n$ values
$\ket{X}=\{x_i\}_{i=1}^n$, where $x_i\in [0,1]$. The quantum system encoding 
the 
data from the feature space will be a direct sum 
$\mathcal{H}=(\mathbb{C}^k)^{\oplus n}$. Quantum 
representation $\ket{\Phi(X)}\in\mathcal{H}$ of a picture $\ket{X}$ will be a 
mapping from $[0,1]^n$ to $\mathcal{H}$ defined by
\begin{equation} 
\ket{\Phi(X)}=\frac{1}{\sqrt{n}}\bigoplus\limits_{i=1}\limits^n 
\ket{\phi(x_i)},
\label{eq:image_representation}
\end{equation}
where pixels are represented by
\begin{equation}
\ket{\phi (x_i)}=x_i\ket{0}+\sqrt{1-x_i^2}\ket{1}.
\end{equation}

We will use quantum representation of vectors from PCA in the same way. Let 
$\{\ket{V_l}\}_{l=1}^s$be a set of principal components with values $v_{l,i}\in 
[-1,1]$ and $\{\ket{\Phi(V_l)}\}_{l=1}^s$ be a set of quantum representations 
of them.
Representation of each quantum vector chosen from PCA is encoded on different 
2-dimensional
subspace of $\mathbb{C}^k$ such that each of the pixels is transformed into
\begin{equation}
\ket{\phi{(v_{j,i})}}=v_{l,i}\ket{0}+\sqrt{1-v_{l,i}^2}\ket{j+1},
\end{equation}
where $j\in \{1,2,3, \ldots, s\}$ and $i$ is a pixel index. The whole 
principal component representation is composed of the pixel representations the 
same way as in eq. (\ref{eq:image_representation}).  Lets take two 
vectors
$\ket{V_{j}},\ \ket{V_{l}}$ and their quantum representation $\ket{\Phi(V_j)},\
\ket{\Phi(V_l)}$. Inner product of $\ket{V_l}$ and $\ket{V_j}$ is
\begin{equation}
\braket{V_j}{V_l}=\sum\limits_{i=1}\limits^nv_{l,i}^*v_{j,i}, 
\label{classic_inner}
\end{equation}
and for corresponding quantum representations one reads
\begin{equation}
\begin{split}
\braket{\Phi(V_j)}{\Phi(V_l)}
&=\frac{1}{\sqrt{n}}\frac{1}{\sqrt{n}}
\sum\limits_{i=1}\limits^n\braket{\phi(v_{j,i})}{\phi(v_{l,i})}\\
&=\frac{1}{n}\sum\limits_{i=1}\limits^n(v_{l,i}^*v_{j,i}\braket{0}{0}+\sqrt{1-v_{j,i}^2}v_{l,i}\braket{j+1}{0}+\\
&\phantom{\ \ }+v_{j,i}^*\sqrt{1-v_{l,i}^2}\braket{0}{l+1}+\sqrt{1-v_{l,i}^2}\sqrt{1-v_{j,i}^2}\braket{j+1}{l+1})\\
&=\frac{1}{n}\sum\limits_{i=1}\limits^nv_{l,i}^*v_{j,i},
\end{split} \label{quantum_inner}
\end{equation}
where the first equality is from eq. (\ref{eq:image_representation}) and the 
last one is implied by orthonormality of the basis vectors.
From eq. (\ref{classic_inner}) and eq. (\ref{quantum_inner}) we derive that inner product of two vectors is equal to inner product of quantum representations of these vectors with respect to a constant factor $1/n $. This is an important feature of the introduced representation, which is significant for the algorithm.
\subsection{Construction of measurement}
The quantum algorithm for principal component analysis is based on classical 
methods for determining the characteristic subspace of the data set in the 
features space.
Thus we take $s$ principal components $\{\ket{V_l}\}_{l=1}^s$ that describe the 
data set crucial properties. 
The quantum algorithm for principal components analysis will utilise the system 
$\mathcal{H}$ defined in the previous section.
In order to use the classically computed 
components in the quantum algorithm we need to convert our principal components 
into quantum representation $\{\ket{\Phi{(V_l)}}\}_{l=1}^s$.

The developed algorithm is based on the quantum measurement schema. 
We consider two elements output set $\Gamma=\{\mathrm{yes}, \mathrm{no}\}$. The first of the resulting 
labels will correspond to the principal components subspace and the other to 
the rest of the feature space. Thus we create two measurement operators $\Pi$ 
and $\mathbb{1}-\Pi$.
The principal components projection operator $\Pi$ is of the form
\begin{equation}
\Pi=\sum\limits_{l=1}\limits^s\ketbra{\Phi(V_l)}{\Phi(V_l)}. \label{proj}
\end{equation}

\subsection{Measurement probabilities}
Let $\ket{X}$ be a input feature vector and $\{\ket{V_l}\}_{l=1}^s$ set of the 
principal components.
In the classical model we measure the likelihood of the input data being in the 
control set in the following way
\begin{equation}
M=\sum_{l=1}^{s}|\braket{V_l}{X}|^2.
\end{equation}
Now let $\ket{\Phi(X)}$ be a quantum representation of the input and 
$\{\ket{\Phi(V_l)}\}_{l=1}^s$ be quantum representation of principal components 
with projector $\Pi$ constructed as in eq. (\ref{proj}). Then the probability
of the result of the measurement being $Y$ for a given input is
\begin{equation}
P_{\Gamma }(\mathrm{yes}|X)=\bra{\Phi(X)}\Pi\ket{\Phi(X)}=\frac{1}{n^2}\sum\limits_{l=1}^{s}|\braket{\Phi(V_l)}{\Phi(X)}|^2=\frac{1}{n^2}M,
\end{equation}
where the last equation results from eq. (\ref{classic_inner}) and eq.
(\ref{quantum_inner}). Thus the probability $P_{\Gamma }(\mathrm{yes}|X)$ is linearly dependent on 
the classical likelihood measure $M$ with respect to a factor $1/n^2$.
where last equation is from eq. (\ref{classic_inner}) and eq.
(\ref{quantum_inner}). Thus the probability $P_{\Gamma }(\mathrm{yes}|X)$ linearly dependent on 
the classical likelihood measure $M$ with factor $1/n^2$.

Because of the factor $1/n^2$ we perform $n^2$ tests.
We assume that we have $n^2$ copies of the quantum representation of the vector 
$\ket{X}$. We perform the measurement $\Pi$ on each of the copies.
If any of the measurements returns ``yes'' then the algorithm returns positive 
answer. If not, the answer is negative.
The probability that our algorithm will return the output ``no''for a given 
input vector $\ket{X}$ is equal to
\begin{equation}
P_{\Gamma, n^2}(\mathrm{no}|X)=(1-P_{\Gamma }(\mathrm{no}|X))^{n^2}.
\end{equation} 
Probability of positively classifying the input image in most of the cases is close to the classical likelihood measure. In general the probability is slightly lower. Thus the algorithm trifle favors the negative answer.
\section{Concluding remarks}
In this paper we provided a new quantum representation of digital images and an 
algorithm for classification of said images. The principal component analysis 
is used during the learning phase of the algorithm which is performed 
classically and its goal is to construct a quantum measuring device.
Classification is performed by applying the measurement apparatus on the 
quantum states that represent input images. The measurement is performed on 
multiple copies of the image. Therefore the paper provides complete system for 
classification of digital images.

\section{Acknowledgements}
Work by M.O. and P.S. was supported by Polish National Science Centre grant 
number DEC-2011/03/D/ST6/00413.
Work by P.G. was supported by Polish National Science Centre grant number
DEC-2011/03/D/ST6/03753.
 
\bibliographystyle{splncs03}
\bibliography{eigenfaces}
\end{document}